\documentclass[preprint2]{aastex}

\usepackage{graphicx}
\usepackage{epsf}

\shorttitle{Photoelectric heating and [CII] cooling}
\shortauthors{Juvela, Padoan \& Jimenez}

\begin{document}

\title{Photoelectric heating and [CII] cooling in translucent clouds: \\
results for cloud models based on simulations of compressible MHD turbulence}

\author{M.~Juvela}
\affil{Helsinki University Observatory,
P.O.Box 14,
SF-00014 University of Helsinki, Finland}
\email{mika.juvela@astro.helsinki.fi}

\and 

\author{P.~Padoan}
\affil{Jet Propulsion Laboratory, 4800 Oak Grove Drive, MS 169-506, California
Institute of Technology}

\and

\author{R.~Jimenez}
\affil{Department of Physics and Astronomy, Univ. of Pennsylvania, Philadelphia,
PA 19104, USA}

\begin{abstract}
Far--ultraviolet (FUV) photons expel electrons from interstellar dust grains
and the excess kinetic energy of the electrons is converted into gas thermal 
energy through collisions. This photoelectric heating is believed to be the 
main heating mechanism in cool HI clouds. The heating rate cannot be directly 
measured, but it can be estimated through observations of the [CII] line emission, 
since this is believed to be the main coolant in regions where the photoelectric 
effect dominates the heating. Furthermore, the comparison of the [CII] 
emission with the far--infrared (FIR) emission allows to constrain the efficiency
of the photoelectric heating, using model calculations that take into account
the strength of the radiation field. Recent [CII] observations carried 
out with the ISO satellite have made this study possible.

In this work we study the correlation between FUV absorption and FIR emission 
using three--dimensional models of the density distribution in HI clouds. The 
density distributions are obtained as the result of numerical simulations 
of compressible magneto--hydrodynamic turbulence, with rms sonic Mach numbers
of the flow ranging from subsonic to highly supersonic, $0.6 \le M_{\rm S}\le 10$. 
The FIR intensities are solved with detailed radiative transfer calculations. 
The [CII] line radiation is estimated assuming local thermodynamic equilibrium where 
the [CII] line cooling equals the FUV absorption multiplied by the unknown 
efficiency of the photoelectric heating, $\epsilon$. 

The average ratio between the predicted [CII] and FIR emissions is found to be 
remarkably constant between different models, implying that the derived
values of $\epsilon$ should not depend on the rms Mach number of the turbulence.
The comparison of the models with the empirical data from translucent, high 
latitude clouds yields an estimate of the photoelectric heating efficiency of
$\epsilon \sim$2.9$\times 10^{-2}$, based on the dust model of Li \& Draine.
This value confirms previous theoretical predictions. 

The observed correlation between [CII] and FIR emission contains a large scatter,
even within individual clouds. Our models show that most of the scatter can be 
understood as resulting from the highly fragmented density field in turbulent
HI clouds. The scatter can be reproduced with density distributions from 
supersonic turbulence, while subsonic turbulence fails to generate the observed 
scatter.

\end{abstract}

\keywords{ISM: clouds --- radio lines: ISM --- infrared: ISM 
--- radiative transfer --- ISM: dust --- ISM: kinematics and dynamics
}

\section{Introduction}

In diffuse interstellar clouds ($A_{\rm V}\la 1$) the gas is heated mainly by
photoelectrons expelled from dust grains by far--ultraviolet (FUV) photons
\citep{jong77,draine78,bakes94,wd134} 
in
the energy range 6\,eV$<h\nu<$13.6\,eV. The high energy limit corresponds to
the cut--off in the FUV radiation field caused by the hydrogen absorption
($h\nu$=13.6\,eV), while the low energy limit corresponds to the energy needed
to free electrons from the grains ($h\nu \sim$6\,eV). In the cold neutral
medium ($T_{\rm kin}\la200$\,K) photoelectric heating accounts for most of the
heating, the X-ray and cosmic ray heating rates being more than an order of
magnitude smaller \citep{wolfire95}. In a relatively dense
neutral medium ($n_{\rm H}\ga$100\,cm$^{-3}$), where a significant fraction of
carbon is in the neutral form, carbon ionization becomes an important heating
source, but it is still not comparable to the photoelectric effect.

The efficiency of the photoelectric heating, $\epsilon$, is defined as the
ratio between the kinetic energy of photoelectrons available to heat the gas
and the FUV energy absorbed by the grains. \citet{bakes94} estimated the
photoelectric yield to be close to 10\%, meaning that of all the photons
absorbed by grains only one in ten ionizes them. Furthermore, since the
average absorbed photon has an energy of $\sim$8\,eV and the work function for
a neutral grain is $\sim$5.5\,eV, only one third of the absorbed energy goes
into the kinetic energy of the electron. The overall efficiency is therefore
expected to be close to 3\%, at least in the cold neutral medium. The
rate decreases as the grains become positively charged. In the intense
FUV radiation field of photodissociation regions, for example, the efficiency
can be more than a factor of ten lower than in the cold neutral medium.

\citet{ingalls02} have recently studied a sample of translucent, high latitude 
clouds (HLCs). The [C$^{\rm +}$] 158\,$\mu$m line $^2P_{3/2}\rightarrow^2P_{1/2}$ 
was observed with ISO and this was correlated with FIR intensities. In the cold 
neutral medium the [CII] line dominates the cooling \citep{wolfire95}, neutral 
carbon being the next most important coolant. \citet{ingalls02} estimated that for
these clouds the CI cooling rate is at most 30\% of the [CII] cooling rate and the
CO rate is more than one order of magnitude smaller. Therefore, for the observed 
HLCs the [C$^{\rm +}$] 158$\mu$m line should account for at least 60\% of the
cooling and in most lines of sight the percentage should be considerably higher.

To be a tracer of photoelectric heating the [CII] emission must originate from
the same regions where the photoelectric effect is the main heating source and,
conversely, all regions where the photoelectric heating is dominant must be cooled
predominantly by the [CII] emission. As discussed by Ingalls et al. (2002), both 
conditions are approximately satisfied. The ionization potential of carbon, 11.3\,eV, 
is above the work function of grains, and therefore C$^{\rm +}$ exists only in regions 
where photoelectric heating is possible. Conversely, in regions where C$^{\rm +}$ is 
not present and C and CO are significant coolants, the photoelectric heating rate 
must be much lower than in regions where C$^{\rm +}$ is present. As in Ingalls et al.
(2002) we therefore assume that the above conditions are satisfied and the
photoelectric heating is exactly balanced by the [CII] line cooling.

Since all the FUV radiation in the energy range 6\,eV$<h\nu<$13.6\,eV absorbed
by grains contributes to the photoelectric heating, the efficiency $\epsilon$
of this heating mechanism can be computed as the ratio of the observed [CII]
line emission and the absorbed FUV radiation. 
If dust properties are assumed to be known, the absorbed FUV radiation can be
estimated through the observed FIR emission.
The problem of estimating $\epsilon$ is therefore reduced to the problem of
computing the correlation between the absorbed FUV radiation and the
observable FIR emission. Such correlation depends on both the intensity of the
radiation field and the column density and can be determined with model
calculations. The degeneracy between the effect of the column density and of
the intensity of the the external radiation field could in principle be broken
using the information contained in the shape of the spectral energy distribution. 
This requires, however, quite precise modeling of the dust emission.

\citet{ingalls02} modeled their observations using a plane
parallel geometry. The FIR emission was calculated with a dust model consisting
only of grains at equilibrium temperature. A comparison with [CII] and FIR
observations resulted in an estimate of $\epsilon=0.043$ for the efficiency of
the photoelectric heating. The average radiation field, $\chi$, was estimated
to be above the local interstellar radiation field (ISRF), $\chi \sim 1.6
\chi_0$, where $\chi_0$ refers to the sum of the 2.7\,K cosmic background and
the spectrum given by \citet{mezger82} and \citet{mathis83}.

In the present work we follow the same approach as in Ingalls et al. (2002),
in the sense of deriving theoretical relations between the FUV absorption 
and the FIR emission and then using the empirical relation between [CII] and FIR 
intensities to estimate the average efficiency of the photoelectric heating.
However, our models of the FUV absorption and FIR emission is significantly
more detailed and realistic than in Ingalls et al. (2002). We compute the radiative
transfer on cloud models based on three dimensional numerical simulations of
compressible magneto--hydrodynamic turbulence. The density distribution of these
turbulent flows provides a realistic model for the density inhomogeneity of
interstellar clouds
\citep{pjn97,pnj97,padoan98,padoan99,Padoan+Nordlund99MHD}.
The penetration of FUV radiation into the cloud depends on the cloud
structure.
Inside an inhomogeneous cloud the intensity of short wavelength radiation is
higher than inside a homogenous cloud. Furthermore, density variations generate 
a significant scatter in the relation between the local FUV absorption and 
FIR emission. Our model can therefore be used to estimate what fraction of the 
observed scatter can be attributed to the inhomogeneous nature of the density 
field. Other factors that could also contribute to the observed scatter, such 
as anisotropy in the radiation field or abundance variations, are not considered. 
Other improvements of our work, compared with Ingalls et al. (2002), are related 
to the employed dust model. We use the three component model of \citet{li01}. The FIR 
intensities are calculated using the method of \citet{juvela03} and the emission from
transiently heated dust grains is included.

\section{The cloud models}

The density distributions of the models are the result of three dimensional
simulations of super--Alfv\'{e}nic, compressible, magneto--hydrodynamic (MHD)
turbulence. Three models are used, from three simulations with different
values of the rms sonic Mach number of the flow, $M_{\rm S}=0.6$, 2.5 and 10.0
in model $A$, $B$, and $C$, respectively.

The simulations are carried out on a staggered grid of 250$^{3}$ computational
cells, with periodic boundary conditions. Turbulence is set up as an initial
large scale random and solenoidal velocity field (generated in Fourier space
with power only in the range of wavenumbers $1\le K\le 2$) and maintained with
an external large scale random and solenoidal force, correlated at the largest
scale turn--over time. The initial density and magnetic field are uniform and
the gas is assumed to be isothermal. Details about the numerical method are
given in Padoan \& Nordlund (1999).

Experiments are run for approximately 10 dynamical times in order to achieve a
statistically relaxed state. The cloud models used in this work correspond to
the final snapshot of each simulation. The value of $M_{\rm S}$ is varied in
different experiments by varying the thermal energy. The initial rms
Alfv\'{e}nic Mach number remains unchanged, $M_{\rm A}=10.0$, from run to run.
The volume--averaged magnetic field strength is constant in time because of
the imposed flux conservation. The magnetic energy is instead amplified. The
initial value of the ratio of average magnetic and dynamic pressures is
$\langle P_{\rm m} \rangle _{\rm in} / \langle P_{\rm d} \rangle _{\rm
in}=0.005$ for all runs, so all the runs are initially super--Alfv\'{e}nic.
The value of the same ratio at later times is larger, due to the magnetic
energy amplification, but still significantly lower than unity (0.21, 0.14 and
0.12 for models $A$, $B$ and $C$ respectively). The turbulence is therefore
super--Alfv\'{e}nic at all times.

Supersonic and super--Alfv\'{e}nic turbulence of an isothermal gas generates a 
highly inhomogeneous gas density distribution, with a density contrast of several 
orders of magnitude. It has been shown to provide a good description of the dynamics 
of molecular clouds and of their highly fragmented nature
\citep{pjn97, pnj97, padoan98, padoan99, Padoan+Nordlund99MHD}.
Translucent high latitude clouds (HLCs), such as the ones observed by
\citet{ingalls02}, have been studied thanks to their CO emission \citep{magnani96}. 
Despite their lower gas density and presumably larger fraction of gas in the
atomic form, the clouds can still be approximately modeled with an isothermal
equation of state.
At densities above 1\,cm$^{-3}$ the gas temperature is expected to remain 
mostly in the range 50--100\,K \citep[e.g.][]{wolfire95}.
The observed clouds are characterized by supersonic turbulent motions like
denser molecular clouds \citep[see e.g.][]{miesch99}.
It seems therefore appropriate to use supersonic turbulence as a model not
only of dense molecular clouds, but also of more diffuse clouds, such as the
ones studied in this work. A subsonic model (model $A$) is also used in this
work for the purpose of studying the effect of varying $M_{\rm S}$ over a
large range of values.

To compute the radiative transfer the data cubes of the density distributions
are scaled to physical units by fixing the length of the computational grid,
$L=6.0$~pc, and the average hydrogen density, $\langle n _{\rm H}\rangle
=100$~cm$^{-3}$. However, the radiative transfer calculations apply to all
values of $L$ and $\langle n _{\rm H}\rangle$ satisfying the condition that
their product is constant, $L\,\langle n _{\rm H}\rangle=600$~pc~cm$^{-3}$=
$1.8\times 10^{21}$~cm$^{-2}$. The column density in the models varies between 
different lines of sight, because of the inhomogeneous density field. Variations 
are particularly large in model $C$, where the range of column density values 
is comparable to that found in the \citet{ingalls02} sample. This can be seen
for example by comparing the FIR intensities in the models and in the observations 
(see Sect.~\ref{sect:results}). 
When results are compared with observations, the angular size of the
models (or their distance) must be fixed as well, in order to take into account
the spatial resolution of the observations. One cell, 6.0/250~pc, is assumed to
correspond to 0.25$\arcmin$, setting the model clouds at the distance of
330~pc.

\section{Calculation of FUV absorption and FIR emission}

Radiative transfer calculations are needed to obtain estimates for the [CII]
emission and the total FIR emission along different lines of sight through the
model clouds. Calculations are carried out on the original grid of 250$^3$
cells. The external radiation field is assumed to be the local interstellar
radiation field near the Sun \citep{mathis83}. The dust model is taken from
\citet{li01}. This includes silicate grains (sizes $a>3.5$~\AA), graphite
grains ($a\ga50$~\AA) and PAHs (from $a=3.5$~\AA\,  to $\sim 50$~\AA) with
ionization state corresponding to physical conditions in the cold neutral medium
\citep[see also][]{juvela03}.

Following \citet{ingalls02} we assume explicitly a balance 
between the cooling by [CII] line emission and the photoelectric
heating. The latter is taken to be equal to the dust absorption in the energy
range 6--13.6\,eV multiplied by the efficiency of the photoelectric emission,
$\epsilon$.
The penetration of the external radiation field is simulated with Monte Carlo
methods, and the absorption in the given energy range is stored in each cell.
The emerging [CII] intensity is calculated by summing the corresponding
emission on selected lines--of--sight, assuming the emission is optically thin.
The efficiency of the photoelectric heating is unknown but since it is assumed
to be constant troughout the cloud it can be taken into account later in
the analysis.

The dust emission was calculated using the `library method' discussed in
\citet{juvela03}. First a mapping between the strength of the
local radiation field at a few reference frequencies and the resulting dust
emission is established. The radiation field at the reference frequencies is
then simulated and the emission is obtained using the existing mapping. The
emission, including the contribution from transiently heated particles, can be
estimated with an accuracy of a few per cent even in the case of large
three--dimensional models \citep[see][]{juvela03}. The infrared emission from
the clouds is estimated using this method.
Ingalls et al. (2002) calculated FIR emission based on 60$\mu$m and
100$\mu$m values using the formula
\begin{equation}
I_{\rm FIR} = 12.6 \,
\frac{I(100\mu m)-2.58\, I(60\mu m)}{\rm MJy\, sr^{-1}}
\times 10^{-6}
\frac{\rm erg}{\rm cm^2\,s\,sr}.
\end{equation}
The formula gives the FIR emission between 42.5\,$\mu$m and 122.5\,$\mu$m i.e.
the wavelength range covered by the IRAS 60$\mu$m and 100$\mu$m filters
\citep{helou85}. In our models the formula was found to be accurate to
$\sim2-3$ per cent. However, the quoted FIR intensities are here calculated by
direct integration over the computed spectra.

Both the FUV absorption (equivalent to the [CII] emission divided by the
efficiency, $I_{\rm CII}/\epsilon$) and the FIR intensity, $I_{\rm FIR}$, are
calculated towards three directions perpendicular to the faces of the cubic
cloud. For each direction, maps of 250$\times$250 positions are generated.
Results are compared with observations presented by 
\citet{ingalls02}. The ISO LWS observations of the [C$^{\rm +}$] 158$\mu$m
line have a resolution of 71$\arcsec$. The FIR data are based on IRAS ISSA
maps with a resolution of $4-5\,\arcmin$.
We assume the cells of our models have an angular size of 0.25$\arcmin$, and
the computed [CII] and FIR maps are convolved to the resolution of the
observations.

\section{Results}  \label{sect:results}

\subsection{FIR and [CII] emission in model clouds} \label{sect:fir_cii}

For each model the FUV absorption and the FIR intensity, $I_{\rm FIR}$ are
calculated towards three directions perpendicular to the faces of the density
field data cubes. 
Since the FUV absorption is assumed to be equivalent to the [CII]
line intensity, $I_{\rm CII}$, divided by the efficiency, $\epsilon$, in the
following we will refer to the quantity $I_{\rm CII}/\epsilon$ (with $\epsilon$
still unknown), instead of the computed FUV absorption.

\begin{figure}
\epsscale{1.0}
\plotone{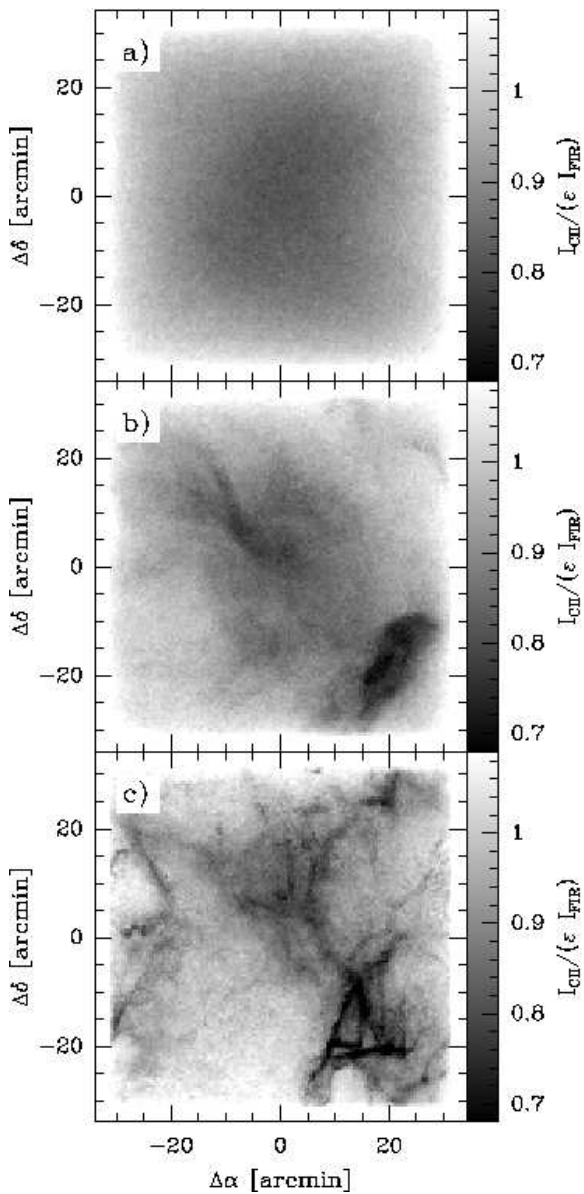}
\caption[]{The ratio $I_{\rm CII}/(\epsilon I_{\rm FIR})$
in the three model clouds $A$, $B$ and $C$ (from top to bottom). 
The maps correspond to observations towards the $y$--direction. 
No convolution has been applied to either $I_{\rm CII}/\epsilon$ or 
$I_{\rm FIR}$ data.
}
\label{fig:map.y}
\end{figure}

Fig.~\ref{fig:map.y} shows the distribution of the ratio $I_{\rm
CII}/(\epsilon I_{\rm FIR})$ in the three models. The figure is obtained at
the original resolution (250$\times$250 pixels), and neither $I_{\rm FIR}$
nor $I_{\rm CII}/\epsilon$ are spatially convolved.
Model $A$ is based on a flow with subsonic turbulence, with only small density 
fluctuations. This is reflected in the top panel of Fig.~\ref{fig:map.y} showing
a smooth distribution of $I_{\rm CII}/(\epsilon I_{\rm FIR})$ in model $A$. 
The density contrast increases with increasing $M_{\rm S}$. Model $B$, with
$M_{\rm S}=2.5$, shows a more `clumpy' distribution of
$I_{\rm CII}/(\epsilon I_{\rm FIR})$ (middle panel of Fig.~\ref{fig:map.y}).
Model $C$, with $M_{\rm S}=10$, has clearly the most inhomogeneous and
filamentary distribution (bottom panel of Fig.~\ref{fig:map.y}).

\begin{figure*}
\epsscale{2.0}
\centering
\plotone{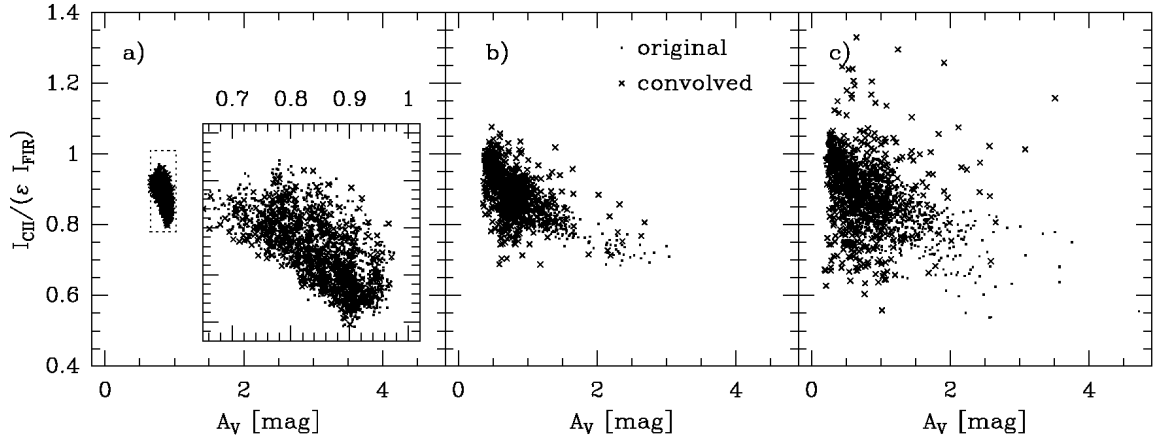}
\caption{
The ratio $I_{\rm CII}/(\epsilon I_{\rm FIR})$ as a function of the visual
extinction along the line of sight. The three panels, a, b and c, correspond 
to the three models, $A$, $B$ and $C$ respectively. The figures show values
calculated from the models towards the $y$--direction. The relations are shown 
for both the original data (dots) and the convolved data (crosses). For 
model $A$ the indicated area has been plotted also on a larger scale.}
\label{fig:av.y}
\end{figure*}

In the case of the nearly homogeneous cloud $A$, both $I_{\rm FIR}$ and
$I_{\rm CII}/\epsilon$ are determined mainly by the distance to the cloud surface. 
Since $I_{\rm CII}/\epsilon$ is assumed to be equivalent to the FUV
absorption, $I_{\rm CII}/\epsilon$ depends only on photons in the
energy range 6--13.6~eV, while $I_{\rm FIR}$ depends on a much wider wavelength 
range responsible for heating the dust grains. The dust extinction
increases towards shorter wavelengths, and this leads to a reduction in the
ratio $I_{\rm CII}/(\epsilon I_{\rm FIR})$ inside the clouds. At the cloud
surface the ratio $I_{\rm CII}/(\epsilon I_{\rm FIR})$ is above one, and it
decreases to $\la$0.8 inside the cloud, in all three models. 
The total range of values is wider in the more inhomogeneous clouds, with the 
lowest values reached in the highest density peaks. The average ratios are,
however, very similar in all three models, $\langle I_{\rm CII}/(\epsilon
I_{\rm FIR}) \rangle \approx$0.9. 
The average value depends, of course, on the total column density and it would 
become lower in more opaque clouds.
The models have similar column densities as the clouds in the sample of
\citet{ingalls02}, and if the clouds are subjected to similar
radiation field as the models the exact correspondence can be checked by
comparing the FIR intensities.

In Fig.~\ref{fig:av.y} we plot the ratio $I_{\rm CII}/(\epsilon I_{\rm FIR})$
against the visual extinction of the corresponding line of sight. The average
optical depth of the model clouds is below one, but the total range of $A_{\rm
V}$ values depends on the degree of inhomogeneity. While in model $A$ the
visual extinction stays below $\sim$1$^{\rm m}$, the other models include a
number of lines of sight with $A_{\rm V}$ in excess of 2$^{\rm m}$ (model $B$)
and 3$^{\rm m}$ (model $C$). The figure shows the
results for both unconvolved data and for the case where data have been
convolved to the resolution of the observations (71$\arcsec$ for [CII] and
$\sim$4$\arcmin$ for FIR data). The $A_{\rm V}$ data has not been convolved. This
would correspond to a situation where the extinction is derived from 
photometry of an individual background star, probing only a single line of 
sight inside the larger beam employed in the other observations.

The negative correlation between extinction and the ratio $I_{\rm
CII}/(\epsilon I_{\rm FIR})$ is clearly visible in model $A$
(Fig.~\ref{fig:av.y}a). In model $B$ the correlation can be seen in the
unconvolved data, while in the convolved data the large difference between the
resolution of the [CII] and FIR observations increases the scatter and the
correlation becomes weak. At high extinction values the convolution increases
systematically the $I_{\rm CII}/(\epsilon I_{\rm FIR})$ values. The [CII]
intensity depends on FUV absorption, and because of the high optical depth the
[CII] emission is likely to saturate near the column density maxima. Since the
intensity distribution is already flat the convolution has little effect on
the peak [CII] intensities. The FIR emission is less saturated and, more
importantly, the FIR intensities are convolved with a beam that is several
times larger. As the the FIR peaks are smoothed away the ratio $I_{\rm
CII}/(\epsilon I_{\rm FIR})$ increases.
The effect is even higher in model $C$ because of the stronger
density contrast. However, in model $C$ the $A_{\rm V}$--dependence is quite
weak even in the unconvolved data. This can be attributed to other effects
related to the increased density contrast. For example, dense filaments and
sheets are common, with the result that the $A_{\rm V}$ along one
line of sight can be much higher than the average extinction by which the
material is shielded towards other directions. In the figure this results in a
larger scatter along the $A_{\rm V}$ axis.

\begin{figure*}
\plotone{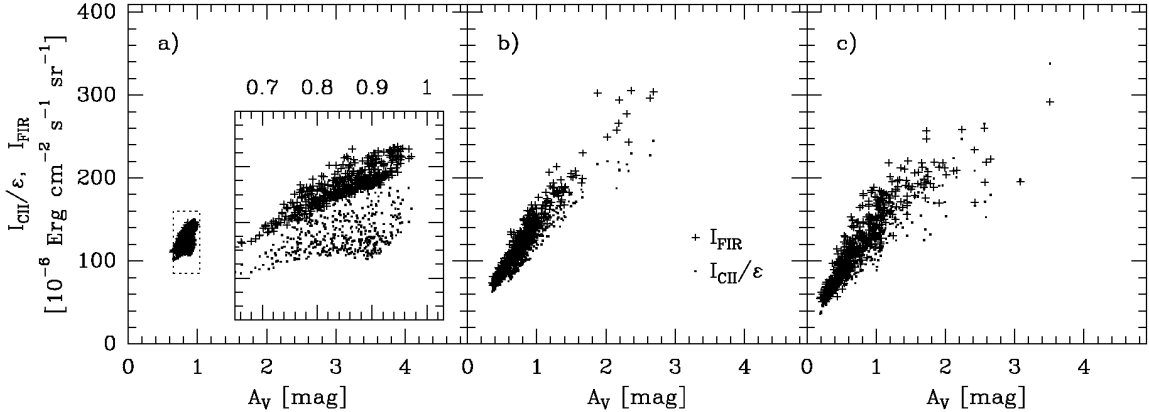}
\caption[]{
The convolved quantities $I_{\rm CII}/\epsilon$ and $I_{\rm FIR}$ as a
function of visual extinction. The three panels, a, b and c, correspond 
to the three models, $A$, $B$ and $C$ respectively. For 
model $A$ the indicated area has been plotted also on a larger scale.
}
\label{fig:av2.y}
\end{figure*}

In a homogeneous model the extinction of the external radiation field proceeds
in a predictable fashion, and the ratio between optical and FUV intensities
only depends on the distance from the cloud surface. In an inhomogeneous cloud
there is no longer a one--to--one relation between the shape of the spectral
energy distribution and the intensity of the radiation. The FUV radiation
penetrates deeper into the cloud than in the homogeneous model, and the
variation in the ratio between optical and FUV intensities becomes smaller. If
the cloud consisted of optically very thick clumps, the incoming radiation
would have exactly the same spectral energy distribution at the centre of the
cloud as on the surface -- only the intensity would be smaller. MHD models are
somewhere between the two extremes, that is between a homogeneous cloud and a
cloud with optically thick clumps. In different parts of the cloud the
spectrum of the radiation field may be different even for the same `effective
extinction' and this increases the scatter along the $I_{\rm CII}/(\epsilon
I_{\rm FIR})$--axis.

In Fig.~\ref{fig:av2.y} the dependence on the visual extinction along the line
of sight is shown separately for $I_{\rm CII}/\epsilon$ and $I_{\rm FIR}$. 
This shows explicitly the saturation of the [CII] emission, which at high
column densities is caused by the total absorption of incoming FUV intensity. The
[CII] line itself is assumed to be optically thin.
The effect is particularly clear in model $B$. It is present also
in model $C$, but it is partially masked by the scatter resulting from
the more inhomogeneous cloud structure and the effects of the convolution.

Inhomogeneity is expected to lead to a decrease in total energy absorption
\citep[see][]{juvela03} and, therefore, to lower the emitted
intensities. The FIR emission in models $B$ and $C$ is respectively 2\% and
5\% lower than in model $A$. The FUV absorption, or the equivalent quantity
$I_{\rm CII}/\epsilon$, is 3.6\% and 9.0\% lower in models $B$ and $C$
respectively than in model $A$. The larger variation of $I_{\rm CII}/\epsilon$
relative to $I_{\rm FIR}$ is the result of a difference in the optical depths
relevant to the two components. For optically thin radiation the absorption
depends only on the total amount of absorbing material while for optically
thick radiation the absorbed energy is proportional to the area filling
factor. Since the optical depth is higher for the FUV radiation than for the
longer wavelengths heating the grains, the quantity $I_{\rm CII}/\epsilon$ is
more sensitive to the density contrast (filling factor) of the cloud. The
equilibrium dust temperature for large grains, and consequently the FIR
emission, depends on longer wavelengths that are partly optically thin.
Therefore, in more clumpy clouds the reduction in $I_{\rm FIR}$ is more modest
than in $I_{\rm CII}/\epsilon$.
However, the value $I_{\rm CII}/(\epsilon I_{\rm FIR})$ decreases from model
$A$ to model $C$ by less than 2\%. The interpretation of observations is
affected by several other uncertainties, such as the imprecise knowledge of
the dust properties or the strength of the external radiation field (see
Sect.~\ref{sect:discussion}). A variation of 2\% is insignificant compared
with these other uncertainties.

\begin{figure*}
\plotone{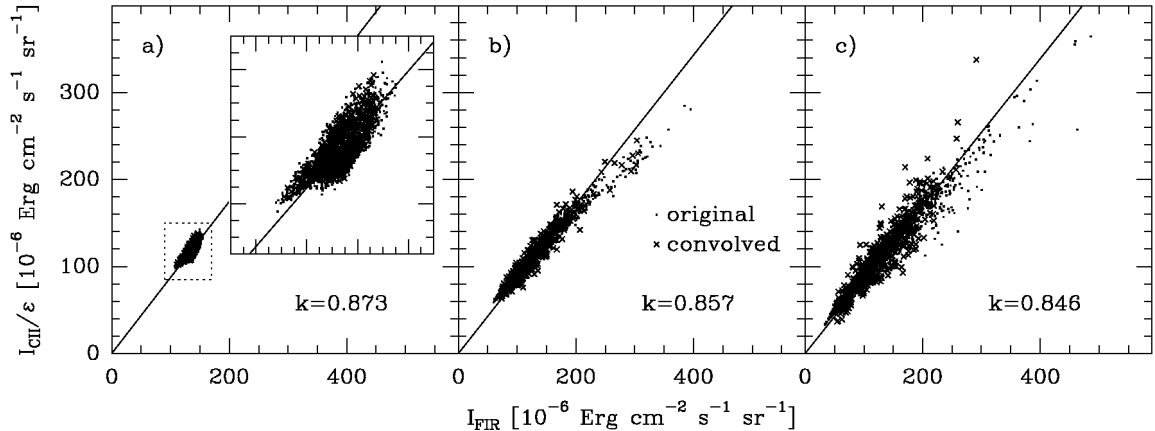}
\caption[]{
$I_{\rm CII}/\epsilon$ vs. $I_{\rm FIR}$ in maps calculated from the three
model clouds towards the $y$--direction. The three panels, a, b and c,
correspond to the three models, $A$, $B$ and $C$ respectively. The [CII] data
has been convolved to a resolution of $\sim$1.2$\arcmin$ and the FIR to a
resolution of $\sim 4\arcmin$. The slope of the least squares line going
through the origin is given in each frame. For model $A$ the data in the
indicated area has been plotted separately on a larger scale.}
\label{fig:cor.y}
\end{figure*}

\subsection{Efficiency of the photoelectric heating} \label{sect:efficiency}

In Fig.~\ref{fig:cor.y} the quantity $I_{\rm CII}/\epsilon$ is plotted versus
the FIR intensity, $I_{\rm FIR}$, for the models $A$, $B$ and $C$ (left to
right) observed along the $y$--direction. We show the relations for both the
original data and the data convolved to the resolution of the observations
discussed in \citet{ingalls02}, that is $\sim 1.2\arcmin$ for [CII]
and $\sim 4\arcmin$ for the FIR observations.
The relations are close to linear, and Fig.~\ref{fig:cor.y} shows the least
squares lines fitted to the convolved data. The range of intensities and the
scatter in the relations both increase from model $A$ to model $C$. The slope
of the relation decreases, albeit very slowly with increasing Mach number of
the model. The fitted slopes for each of the three orthogonal directions of
the lines of sight are given in Table~\ref{table:ratio}.

In Sect~\ref{sect:fir_cii} the [CII] intensities were found to saturate at high
column densities when all incoming FUV radiation was absorbed. As a result, at
high intensities the points for unconvolved data fall below the least squares
line. However, at the location of intensity maxima the convolution decreases
the FIR intensities more than the [CII] intensities, and in the figure the
convolved points are shifted to the left more than down. This is seen especially in model
$C$ that has the largest density contrast. In model $C$ the relation $I_{\rm
CII}/\epsilon$ versus $I_{\rm FIR}$ for the convolved data turns slightly
upwards at higher intensities, relative to the relation for the unconvolved
data. In model $B$ the convolution raises the points at high-$A_{\rm V}$ only
slightly closer to the fitted least squares line.

A comparison between the ratios $I_{\rm CII}/(\epsilon I_{\rm FIR})$ obtained
from the models and the empirical relation $I_{\rm CII}/I_{\rm FIR}$ allows
to estimate the efficiency of the photoelectric heating, $\epsilon$. In
\citet{ingalls02} the average ratio of CII and FIR intensities
is found to be $I_{\rm CII}/I_{\rm FIR}$=(2.5$\pm$0.9)$\times 10^{-2}$, for a
sample of high latitude clouds observed with ISO. The slope they derive for
large scale correlation using DIRBE and FIRAS observations for $|b|>5\deg$ is
also very similar, $I_{\rm CII}$=(2.54$\pm$0.03)$\times 10^{-2} I_{\rm FIR}$ -
(1.1365$\pm$0.0004)$\times 10^{-6}$ erg\,cm$^{-2}$\,s$^{-1}$\,sr$^{-1}$.

\begin{deluxetable}{cccc}
\tablecaption{The slope of the relation $I_{\rm CII}/\epsilon$ versus $I_{\rm FIR}$
in the model clouds observed from three different directions. The values are
based on convolved data. The efficiency of the photoelectric heating,
$\epsilon$, is obtained assuming the empirical relation $I_{\rm CII}/I_{\rm
FIR}$=2.5$\times 10^{-2}$ from Ingalls et al. (2002).}
\tablewidth{0pt}
\tablehead{
Model   &   direction &     $<I_{\rm CII}/(\epsilon I_{\rm FIR})>$
        &   $100\times\epsilon$ }
\startdata
A       &   x         &  0.876    &  2.85    \\
A       &   y         &  0.873    &  2.86    \\
A       &   z         &  0.872    &  2.87    \\
B       &   x         &  0.869    &  2.88    \\
B       &   y         &  0.861    &  2.90    \\
B       &   z         &  0.870    &  2.87    \\
C       &   x         &  0.862    &  2.90    \\
C       &   y         &  0.862    &  2.90    \\
C       &   z         &  0.858    &  2.91    \\
\enddata
\label{table:ratio}
\end{deluxetable}

We have calculated the efficiencies of the photoelectric
emission, $\epsilon$, assuming $I_{\rm CII}/I_{\rm FIR} = 2.5 \times 10^{-2}$. 
The results are given in the fourth column of Table~\ref{table:ratio}.
We obtain an average value of $\epsilon = 2.88 \times 10^{-2}$. There is very 
little variation between the models or the direction from which the model clouds 
are observed. However, as seen in Fig.~\ref{fig:map.y}, the variation within 
individual maps can be quite significant. In the case of unconvolved data in 
Fig.~\ref{fig:map.y} the variation ranges from $\sim$40\% in model $A$ to almost 
90\% in model $C$.

\section{Discussion}  \label{sect:discussion}

The observations presented by \citet{ingalls02} contain
scatter in the plot of $I_{\rm CII}$ versus $I_{\rm FIR}$ exceeding the
uncertainty of individual observations. 
Differences between clouds can be attributed largely to differences in
the dust composition. The photoelectric heating and therefore [CII] emission
depends on the abundance of very small dust grains and PAH--molecules while the
far--infrared emission is caused mainly by large dust grains. The ratio between 
the abundances of very small grains and large grains is known to show
significant spatial variations. The variations are more noticeable in dense clouds 
\citep[e.g.][]{laureijs91,juvela02,stepnik03}, but changes have been
observed even in cirrus type clouds \citep{bernard99,cambresy01}. Variations
in the radiation field are probably less important.

\begin{figure}
\epsscale{1.0}
\plotone{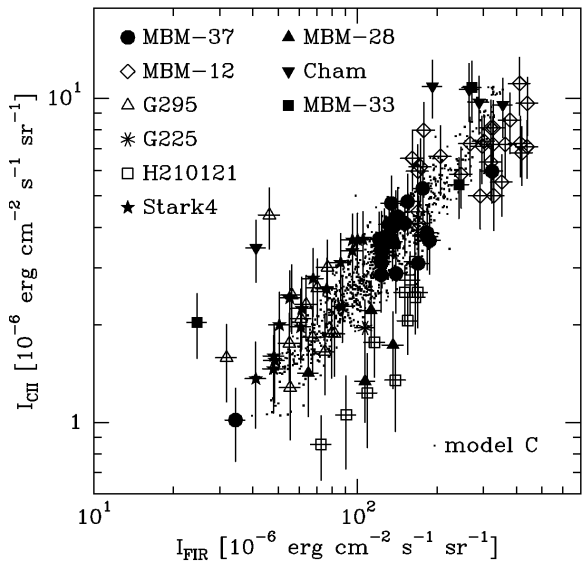}
\caption[]{Comparison of the [CII] and FIR intensities in model $C$ and in the
sample of HLCs in \citet{ingalls02}. For model $C$ the
points correspond to convolved data, and to an efficiency of the
photoelectric heating of 2.88$\times 10^{-2}$
}
\label{fig:ingalls}
\end{figure}

In the first approximation an increase in the strength of the
radiation field affects FIR and [CII] intensities in the same way. However, as
the increased dust temperature moves dust emission to shorter wavelengths the
FIR intensity is reduced relative to the [CII] emission (FUV absorption). 
Assuming that the FIR intensity in the models is fixed and constrained by the 
observations, if the interstellar radiation field is increased the dust column 
density must be decreased to maintain the FIR intensity unchanged. As a result
of the decreased column density, the [CII] emission is less saturated, and the 
ratio $I_{\rm CII} / I_{\rm FIR}$ is increased.
Variations within individual clouds could similarly be caused by variations in
the local radiation field or in the properties of the gas or dust.

In the present models dust properties are assumed to be constant throughout
the clouds. The fractional abundance of C$^{\rm +}$ is also kept constant and the [CII]
emission is assumed to be responsible for all of the gas cooling. Even with
these simplifying assumptions the computed ($I_{\rm CII}$,$I_{\rm FIR}$)
values contain a significant scatter. The scatter is smallest in the most
homogeneous model $A$. Only in a completely homogeneous model the points would
fall on one line as both [CII] and FIR emission would depend solely on the
distance from the cloud surface. The scatter is largest in model $C$, with the
largest value of $M_{\rm S}$ and density contrast. In this model, both the
range of $I_{\rm CII}$ and $I_{\rm FIR}$ values and the scatter in their
relation are similar to the observations presented by Ingalls et al. Such a
comparison is relevant only if the observed positions were selected randomly
within the clouds. This is not always the case, but the scatter of the
observed points should still be close to the true variation within the
observed clouds.

In Fig.~\ref{fig:ingalls} we compare our model $C$ with the observation
presented by \citet{ingalls02}. We have plotted in the same
figure our model $C$ (direction $y$) for which the [CII] intensities have been
calculated using the value $\epsilon=2.9\times 10^{-2}$ derived in
Sect.~\ref{sect:efficiency}. The model covers a large fraction of the observed
range of $I_{\rm CII}$ and $I_{\rm FIR}$ values. After subtracting a linear fit, 
the remaining scatter in all observations is about twice the scatter in model $C$, 
even when the observational errors are taken into account. On the other hand, 
in individual clouds the variation is very similar to that found in the model. 
Our results show that most of the observed scatter may originate from the 
inhomogeneous density structure of the clouds, and the observed scatter is
consistent with that predicted with supersonic and super--Alfv\'{e}nic MHD 
turbulence with sonic rms Mach number $M_{\rm S}\ga2$. In subsonic
models both the scatter and the total range of [CII] and FIR
intensities is instead significantly smaller than in the observed clouds.

Compared with model $C$ the scatter is significantly larger only in the cloud
MBM-12, which also has the highest FIR intensities. MBM-12 has an HI column
density of $\sim$1.1$\times 10^{21}$\,cm$^{-2}$, i.e. some 40\% below the
average column density of the models. \citet{timmermann98} suggested the cloud
is subjected to stronger than average radiation field, $\chi
\sim$1--2\,$\chi_0$, and this could also explain the high FIR intensities. On
the other hand, since the cloud has significant molecular line emission
\citep{magnani96} the total column density may not be very different from the
model, and therefore the observed FIR intensities do not imply a strong
radiation field by themselves. Furthermore, if one would accept the hypothesis
of a strong radiation field, the low FIR colour temperature \citep[see][Fig.
4]{ingalls02} would require a very low abundance of small dust grains.

The scatter in the observed [CII] and FIR intensities may depend on two 
effects not included in the models: i) CI cooling and ii) optical depth 
of the [CII] line. Although the models assume all the cooling is from C$^{\rm +}$, 
CI emission may be important on some lines of sight, increasing the observed scatter. 
According to \citet{ingalls02} the {\em maximum} contribution to the cooling from CI 
is approximately 30\%. The scatter due to the the presence of CI cooling on some
lines of sight should therefore be well below 15\%. Furthermore, the CI contribution 
can be significant only for high column density, and could not explain the
scatter at low FIR intensities. The models also assume that the [CII] line is 
optically thin. If the line became optically thick at some lines of sight,
the main effect would be to flatten the $I_{\rm CII}$ vs. $I_{\rm FIR}$
relation at high column densities, while the scatter would not be significantly
affected. In the models, the saturation of $I_{\rm CII}$ is caused by the total 
absorption of the FUV radiation that is converted into [CII] emission, 
not by self--absorption of the [CII] line itself. Since the observations do not 
show strong [CII] saturation with increasing FIR intensity, the optical depth of 
the [CII] line cannot be very high.

The derived value of the photoelectric heating efficiency, $\epsilon$, is
close to theoretically derived values \citep{jong77,bakes94}. \citet{bakes94}
performed theoretical calculations using a MRN type grain size
distributions \citep[$n\sim a^{-3.5}$,][]{mathis77}. They found that about half
of the heating rate is due to grains with sizes below 15~\AA\, and the
contribution from grains larger than 100~\AA\, is insignificant. The net
efficiency was calculated as a function of the parameter $G_0 (T/{\rm
K})^{1/2} (n_{\rm e}/{\rm cm}^{-3})^{-1}$, 
where $G_0$ is the intensity of the far-UV radiation field relative to
Habing's field \citep{habing68}, 
$T$ is gas temperature and $n_{\rm
e}$ the electron density. The estimated efficiency is constant $\epsilon \sim
3\times 10^{-2}$ for $G_0 (T/{\rm K})^{1/2} (n_{\rm e}/{\rm cm}^{-3})^{-1} \la
10^3$. This prediction should apply to the cloud sample in 
\citet{ingalls02}. Bakes \& Tielens found, however, that the predicted
heating rate could be doubled if the exponent of the grain size distribution
was decreased from -3.5 to -4.0. Since $A_{\rm V}$ and FIR intensity depend on
large grains and the heating mainly on PAHs the derived efficiency $\epsilon$
is model dependent.

In this paper we have used the dust model of \citet{li01}. \citet{wd134} based
their studies on the same dust model and considered further different size
distributions discussed in \citet{wd548}. For cold neutral medium
($T\sim$100\,K) and size distributions consistent with the diffuse cloud
extinction law, $R_{\rm V}\sim$3.1, their predictions for the photoelectric
heating rate are close to the values given by \citet{bakes94}. The results
depend on the carbon abundance which is limited by extinction measurements to
values below $\sim6\times 10^{-5}$ carbon atoms per H nucleus \citep{wd548,
wd134, li01}. The models of \citet{li01} and \citet{wd548} favoured carbon
abundances close to this upper limit and the resulting rate of the
photoelectric heating is some 25\% above the value given by Bakes \& Tielens
\citep{wd134}.

\citet{habart01} derived values of $\epsilon\,$=2--3\% for the
efficiency of the photoelectric heating across the cloud L1721. Compared with
the Ingalls et al. sample, the cloud has higher visual extinction, $A_{\rm
V}\sim3^{m}$, and, due to the proximity of a B2 star, it is subjected to a
stronger radiation field, $\chi\ga5 \chi_0$. By adopting the dust model of
\citet{desert90}, Habart et al. were able to derive
efficiencies of the photoelectric emission separately for PAH, very small
grain and large grain components. The derived efficiencies were $\sim 3$\% for
PAHs, $\sim$1\% for very small grains and $\sim$0.1\% for large grains. Part
of the observed variation was attributed to changes in the abundance of the
dust components. The clouds in the Ingalls sample have lower visual
extinction, and abundance variations are expected to be correspondingly
smaller.

Another source of uncertainty is the balance between the column density and
the strength of the radiation field. As discussed above, if a stronger radiation
field or a smaller column density were adopted in the models, the calculated ratio
$I_{\rm CII}/(\epsilon\,I_{\rm FIR})$ would increase and the derived value
of $\epsilon$ decrease. According to Fig.~\ref{fig:map.y}, the ratio
$I_{\rm CII}/(\epsilon I_{\rm FIR})$ increases up to 30\% above its average
value when moving from the cloud center to the cloud edges, where the
radiation field is not attenuated. Therefore, if the clouds had significantly
lower extinction, the efficiency of the photoelectric heating could be lower
by the same amount.
We modified the model $C$ by reducing its column density by a factor of three
and by multiplying the intensity of the external radiation field by a factor
of two. The estimated average efficiency of the photoelectric heating
decreased from 2.88$\times 10^{-2}$ to 2.25$\times 10^{-2}$, i.e. only by
22\%. 
Based on HI and molecular line data the column densities are in most of
the observed clouds within a factor of two from the average column density of
the models. In this paper we have used for the intensity of the local ISRF
values given by \citet{mezger82} and \citet{mathis83}. According to
\citet{draine78} and \citet{parravano03} the actual ISRF could be up to
$\sim$70\% stronger. However, the difference is very unlikely to be a factor
of two. The actual errors caused by the uncertainty of the column density
and intensity values is therefore less than 22\%.

Compared with \cite{ingalls02} our estimate of the photoelectric heating
efficiency is smaller by one third. Most of the difference is due to a
difference in the predicted FIR intensity. For example, for a model with
optical depth $A_{\rm V}$=1.0~mag through the cloud Ingalls et al. obtained a
100~$\mu$m surface brightness of slightly more than 15\,MJy\,sr$^{-1}$. For a
homogeneous, spherically symmetric cloud with equal optical depth we obtain a
surface brightness of $\sim$9.5\,MJy\,sr$^{-1}$. This is close to what
\citet{bernard92} obtained for a similar cloud model using the
dust model of \citet{desert90}. The difference between our
result and Ingalls et al. cannot be due to model geometry. One would expect
{\em more} emission (higher $I_{\rm FIR}$) from a three dimensional cloud
heated from all directions than from a plane parallel cloud heated on two
surfaces only. 
The most likely explanation for the high FIR intensities obtained by
\citet{ingalls02} is their assumption of thermal equilibrium. This means that
all grains are constantly at temperatures close to 20\,K and almost all
absorbed energy is re-radiated in the FIR. In our case a larger fraction of the
absorbed energy is radiated at shorter wavelengths because of the small
particles that are temporarily heated to much higher temperatures.

\section{Conclusions} \label{sect:conclusions}

We have calculated FIR emission and [CII] line emission for
three--dimensional density distributions of compressible magneto--hydrodynamic
turbulent flows with rms sonic Mach numbers $M_{\rm S}=$~0.6, 2.5, and 10.0. 
The dust emission, $I_{\rm FIR}$, is computed with full radiative transfer 
calculations. The [CII] emission is
estimated assuming the photoelectric heating caused by FUV photons between
0.0912$\mu$m and 0.2066$\mu$m is balanced by [CII] line emission. The FUV
absorption is determined by the radiative transfer simulations, and is
assumed to be equal to the [CII] line intensity divided by the unknown 
efficiency of the photoelectric heating, $I_{\rm CII}/\epsilon$.

The {\em average} ratio $I_{\rm CII}/(\epsilon I_{\rm FIR})$ is in all models
between 0.85 and 0.88, showing that its dependence on the density contrast 
(Mach number) is weak. However, dense filaments are visible in the maps as 
regions with lower value of $I_{\rm CII}/(\epsilon I_{\rm FIR})$. 
The degree of correlation between $I_{\rm CII}/(\epsilon I_{\rm FIR})$ and
visual extinction decreases in more inhomogeneous clouds. In
the case of simulated observations convolved to different resolutions
(71$\arcsec$ for [CII] and $\sim$4$\arcmin$ for FIR) most of the correlation is lost.

The scatter in the observational ($I_{\rm FIR}$,$I_{\rm CII}/\epsilon$) plot
can be reproduced by models with rms Mach number $M_{\rm S}\ga$2 (supersonic turbulence),
showing that most of the scatter may be due to the inhomogeneous nature of the clouds
(likely due to the turbulence). In subsonic models ($M_{\rm S}\la$1) both the scatter 
and the total range of FIR and [CII] intensities become smaller than in the observed
clouds. Using the empirical value $I_{\rm CII}/I_{\rm FIR}$=2.5$\times 10^{-2}$
found for high latitude clouds \citep{ingalls02} the 
efficiency of the photoelectric heating is found to be 
$\epsilon \sim 2.9\times 10^{-2}$. The value is very close to the theoretical 
predictions for the cold neutral medium.

\acknowledgements
MJ acknowledges the support from the Academy of Finland Grants no. 174854 and
175068.
The work of PP was partially performed while PP held a National Research
Council Associateship Award at the Jet Propulsion Laboratory, California
Institute of Technology. The work of RJ is supported in part by NSF grant
AST-0206031.

\end{document}